\documentclass[12pt,twoside]{article}

\usepackage{amsmath, amsthm, amssymb}
\usepackage{graphicx}

\date{}

\begin{document}

\centerline {\Large{\bf Accurate analytical approximation formulae for}}

\centerline{ \Large{\bf large-deviation analysis of rain formation}}


\centerline{}

\newcommand{\mvec}[1]{\mbox{\bfseries\itshape #1}}

\centerline{\bf {Christian P. H. Salas}}


\centerline{Faculty of Mathematics \& Statistics, The Open University, Milton Keynes} 
\centerline{e-mail: c.p.h.salas@open.ac.uk}







\centerline{}\bigskip

\centerline{\bf {Abstract}}\bigskip

\textit{A 2016 paper by M Wilkinson in Physical Review Letters suggests that large-deviation theory is a suitable framework for studying the phenomenon of unexpectedly rapid rain formation in collector-drop collision processes. Wilkinson derives asymptotic approximation formulae for a set of exact large-deviation functions in the collector-drop model, such as the cumulant generating function and the entropy function. The asymptotic approach assumes a large number of water droplet collisions and is motivated by the fact that the exact large-deviation functions are prohibitively difficult to interpret and deal with directly. Wilkinson uses his asymptotic formulae to obtain further results and also provides some numerical work which suggests that a certain log-density function for the collector-drop model (which is a function of his asymptotic approximation formulae) is itself approximated satisfactorily. However, the numerical work does not test the accuracy of the individual asymptotic approximation formulae directly against their exact large-deviation theory counterparts. When these direct checks are carried out, they reveal that the asymptotic formulae are, in fact, rather inaccurate, even for very large numbers of collisions. Their individual inaccuracy is masked by their incorporation into log-density functions in Wilkinson's numerical work. Their inaccuracy, as well as some assumptions underlying their derivation, severely limit their applicability. The purpose of the present note is to point out that it is quite possible to develop accurate \underline{analytical} (i.e., non-asymptotic) approximation formulae for the large-deviation theory functions in the collector-drop model which also preserve the forms of the leading order power terms in Wilkinson's asymptotic formulae. An analytical approximation approach can be developed based on a Euler-Maclaurin formula. The resulting analytical formulae are extremely accurate and valid for all relevant numbers of collisions and time scales, producing numerical results which are essentially indistinguishable from the exact function values of their large-deviation theory counterparts.}\bigskip


\section{Introduction}
M Wilkinson’s 2016 Physical Review Letter \cite{MW2016}, referred to as MW2016 herein, suggests that large-deviation theory is the correct framework for analysing the implications of a collector-drop model of rapid rain formation which had previously been discussed in a paper by Kostinski and Shaw \cite{Kostinski}. A microdroplet in a cloud falls under the influence of gravity and coalesces at a rapidly increasing rate with many smaller water particles. After a certain time and a certain large number of collisions, the droplet enters a stage of \emph{runaway growth} and leaves the system as a raindrop. The collision rates are assumed to increase according to a simple power law function of the number of collisions, with a power law parameter $\gamma$ taking a value around 2 near the start of the process, and a value closer to $4/3$ after the microdroplet reaches a certain size. \newline
\indent In addition to formulating a new volume fraction cut-off mechanism for raindrop size, MW2016 also uses large-deviation theory in conjunction with asymptotic approximations to study the link between the time scale for rapid onset of rain showers and the number of water particle collisions needed for runaway raindrop growth. A Supplement to the main Letter, referred to as MW2016S herein, provides additional details of the (quite intricate) derivations of the asymptotic approximation formulae presented in the main Letter. \newline
\indent The asymptotic approach in MW2016 involves allowing time scales for runaway growth to be variable while assuming a very large number of collisions prior to runaway growth. One problem with this is that Wilkinson's asymptotic formulae are then not directly applicable to other situations which might be of interest requiring \emph{both} the number of collisions and the time scales to be allowed to vary over intermediate values. One might consider making \emph{ad hoc} modifications to the asymptotic formulae to try to make them applicable to more general scenarios, but these efforts would be unlikely to produce numerically accurate results. It would therefore be desirable to try to derive \emph{analytical} (i.e.,  non-asymptotic) formulae instead, accurate for all time scales and numbers of collisions. The main aim of the present note is to show that the development of such analytical approximation formulae is quite feasible. \newline
\indent Another problem with the asymptotic formulae in MW2016 is that they are actually rather inaccurate as approximations of the corresponding exact large-deviation theory functions, even for very large numbers of collisions of order $10^4$ or more. Detailed numerical investigations in which the asymptotic formulae in MW2016 were compared with fine-grained exact function values for the corresponding large-deviation theory functions, both for $\gamma = 2$ and $\gamma = 4/3$, showed that the asymptotic formulae perform poorly. These findings were initially surprising, as the numerical work reported in MW2016 seems to indicate good agreement between approximate and exact \emph{log-density functions} formed from combinations of the relevant asymptotic and exact functions. The individual inaccuracy of the asymptotic formulae in MW2016 is in fact being masked by their incorporation into the log-density functions used for the numerical work in MW2016. It is only when they are compared individually with their large-deviation theory counterparts that their inaccuracy becomes obvious. \newline
\indent In seeking to identify the sources of the inaccuracies in the asymptotic formulae, one is drawn in particular to a sequence of asymptotic simplifications leading to a crucial formula $A(\gamma)$ defined in equation (15) in MW2016S  (see also equation (21) in MW2016). The problem is that these simplifying asymptotic approximations make $A(\gamma)$ independent of certain key variables in the collector-drop model, but it turns out that derivatives with respect to these variables are important for achieving greater approximation accuracy. As described below, a non-asymptotic version of this formula can be obtained which converges to $A(\gamma)$ asymptotically, but which is otherwise allowed to vary with the aforementioned key variables, and this immediately results in greater accuracy. Since $A(\gamma)$ appears as a constant with respect to the key variables in MW2016, the derivatives that are contributing to the much greater accuracy in the non-asymptotic formulation are simply missing in the asymptotic formulae. The approximation errors in MW2016 are further compounded by additional oversimplifications which are quite easily avoided, and remedying these again dramatically improves the overall approximation accuracy. \newline
\indent Section 2 outlines the main mathematical developments in MW2016 leading to the asymptotic formulae, and their inaccuracy is demonstrated via numerical comparisons with corresponding exact large-deviation theory functions. Section 3 sets out the analytical (non-asymptotic) approximation formulae proposed in the present note, gives mathematical details of their derivation, and reports numerical results demonstrating their high degree of accuracy. Section 4 concludes with a brief discussion. 
\section{The asymptotic formulae in MW2016}
In the collector-drop model, a runaway droplet falls and collides with a large number $\mathcal{N}$ of small droplets of radius $a_0$ according to an inhomogeneous Poisson process. If the random variable $t_n$ represents the inter-collision time leading up to the $n$-th collision, the time for a droplet to experience runaway growth is then the random variable 
\begin{equation}
T = \sum_{n=1}^{\mathcal{N}} t_n
\end{equation} 
The inter-collision times $t_n$ are independent exponentially distributed random variables with
\begin{equation}
P_n(t_n) = R_n \exp(-R_n t_n)
\end{equation}
where $R_n$ is the number of collisions per second, i.e., the collision rate, by the time of the $n$-th collision. The collision rate $R_n$ is modelled as 
\begin{equation}
R_n = R_1 f(n)
\end{equation}
where
\begin{equation}
R_1 = \epsilon N_0 \pi(a_0 + a_1)^2 \alpha (a_1^2 - a_0^2)
\end{equation}
is the collision rate of a drop of radius $a_1$ with droplets of radius $a_0$. Here, $\epsilon$ is the coalescence efficiency, $N_0$ is the number density of microscopic water droplets per $m^3$, $\pi(a_0 + a_1)^2$ is the effective cross-section area and $\alpha (a_1^2 - a_0^2)$ is the relative velocity. The constant $\alpha$ is a constant of proportionality between terminal velocity and the radius squared of a water particle. 

The function $f(n)$ in (3) characterises the dependence of $R_n$ on the number of collisions. In \cite{Kostinski}, $f(n) = n^2$, but we can have $f(n) = n^{4/3}$ when $\epsilon \approx 1$ and $a_n \gg a_1$. MW2016 treats $f(n)$ as a simple power law, so that
\begin{equation}
R_n = R_1 n^{\gamma} 
\end{equation} 
The average inter-collision time leading to the $n$-th collision is then given by 
\begin{equation}
\langle t_n \rangle \equiv \tau_n = \frac{1}{R_n} = \frac{1}{R_1} n^{-\gamma}
\end{equation}
and the mean time for explosive growth is obtained as
\begin{equation}
\langle T \rangle = \sum_{n=1}^{\mathcal{N}} \tau_n = \frac{1}{R_1} \sum_{n=1}^{\mathcal{N}} n^{-\gamma}
\end{equation}
When $\gamma > 1$, the mean time for explosive growth converges as $\mathcal{N} \rightarrow \infty$, giving
\begin{equation}
\lim_{\mathcal{N} \rightarrow \infty} \langle T \rangle = \frac{1}{R_1} \zeta(\gamma)
\end{equation}
where $\zeta(\gamma)$ is Riemann's zeta function. 

A cut-off mechanism for raindrop size is obtained by considering the liquid water content of a cloud, expressed as a volume fraction $\Phi_l$. Since $N_0$ is the number of water particles of radius $a_0$ per $m^3$, and each particle is a sphere with volume $\frac{4}{3} \pi a_0^3$, the volume fraction of liquid water per $m^3$ of cloud is given by 
\begin{equation}
\Phi_l \approx N_0 \cdot \frac{4}{3} \pi a_0^3 
\end{equation} 
For example, if each droplet is initially of radius $a_0 = 10 \mu m = 10^{-5}m$, and there are $N_0 \approx 10^9$ such droplets per $m^3$ of cloud, then $\Phi_l \approx 10^{-6}$. 

Using (2), the fraction of droplets undergoing runaway growth between times $t$ and $t + \delta t$ is given by
\begin{equation}
P(t) (t + \delta t - t) = P(t) \delta t
\end{equation} 
For example, if $P(t) \delta t \approx 10^{-6}$, then `one in a million' water droplets will undergo runaway growth between times $t$ and $t + \delta t$. 

If the volume of each runaway droplet increases by a factor of $\mathcal{N}$, the volume fraction of liquid water removed from the cloud per $m^3$ over the time interval of length $\delta t$ will then be 
\begin{equation}
\mathcal{N} \Phi_l P(t) \delta t
\end{equation}

As $\delta t \rightarrow 0$, the instantaneous change of the liquid water content of a cloud per $m^3$ due to runaway growth of droplets is thus
\begin{equation}
\frac{d \Phi_l}{d t} = -\mathcal{N} \Phi_l P(t)
\end{equation}
Therefore, the instantaneous \emph{percentage rate} of decline of the volume fraction is
\begin{equation}
-\frac{d \ln \Phi_l}{d t} = \mathcal{N} P(t)
\end{equation} 
Integrating (13) over a time interval then gives a percentage loss $\mu$ in that interval:
\begin{equation}
\mu = \mathcal{N}\int_0^t dt^{\prime} P(t^{\prime})
\end{equation}

The onset of a shower is determined by the criterion that $\mu$ be a significant percentage of the liquid water content of a cloud that is removed by raindrop formation as a result of $\mathcal{N}$ particle collisions. We are now interested in finding the time scale $t^{*}$ over which this significant reduction in $\Phi_l$ occurs. In accordance with (14), this $t^{*}$ satisfies 
\begin{equation}
\mathcal{N^{*}}\int_0^{t^{*}} dt^{\prime} P(t^{\prime}) = 1
\end{equation}   
where $\mathcal{N^{*}} \equiv \mathcal{N}/\mu$. 

To make progress from here, MW2016 approximates $P(t)$ in (15) using the large-deviation principle, which is explained in detail in, e.g., \cite{Touchette}. With $T$ defined as in (1) and $\langle T \rangle$ given by (7), we implement the large-deviation principle here by approximating the probability $P(T)dT$ at some value $\bar{T}$ in the small-value tail of the distribution of $T$ as
\begin{equation*}
P(\bar{T})d\bar{T} = P(T \in [\bar{T}, \bar{T} + d\bar{T}])
\end{equation*} 
where $P(\bar{T})$ is given by 
\begin{equation}
P(\bar{T}) = \frac{1}{\langle T \rangle} \exp[-J(\tau)]
\end{equation}
with $\tau \equiv \bar{T}/\langle T \rangle$. The function $J(\tau)$ in (16) is referred to as an \emph{entropy function} or \emph{rate function} in large-deviation theory, and is defined below. Using (16) in (15) we get the condition for the onset of rainfall as 
\begin{equation}
\mathcal{N^{*}} \int_0^{t^{*}} \frac{d \bar{T}}{\langle T \rangle} \exp \bigg[-J\bigg(\frac{\bar{T}}{\langle T \rangle}\bigg)\bigg] = \mathcal{N^{*}} \int_0^{t^{*}} d \tau^{\prime} \exp [-J(\tau^{\prime})] = 1 
\end{equation}
with $\tau^{\prime} \equiv \bar{T}/\langle T \rangle$. A saddle-point approximation for the integral here at the maximum value of the integrand, $\exp[-J(\tau^{*})]$, with $\tau^{*} \equiv \bar{T}^{\ast}/\langle T \rangle$, then gives the condition for the onset of rainfall as
\begin{equation}
\mathcal{N}^{*} \exp[-J(\tau^{*})] = 1
\end{equation} 
Taking logs, we can write the condition alternatively as
\begin{equation}
\ln \mathcal{N}^{*} = J(\tau^{*})
\end{equation}
What is required now is to find an expression for $\tau^{*}$ in terms of $\ln \mathcal{N}^{*}$, using (19). To obtain this, it is necessary to find an expression for the rate function $J(\tau)$ for the random sum defined by (1) and (2). 

Consider a single inter-collision time $t_n$ with density (2). Following \cite{Touchette} for the purposes of the large-deviation analysis here, we obtain the log of the Laplace transform of the random variable $t_n$, often referred to as its \emph{scaled cumulant generating function}, as 

\begin{equation}
\lambda_n (k) = -\ln \langle \text{e}^{-kt_n} \rangle = -\ln\bigg(\int_0^{\infty} d t_n \text{e}^{-kt_n} R_n \text{e}^{-R_n t_n}\bigg) = \ln\bigg(1 + \frac{k}{R_n}\bigg)
\end{equation}
To ensure the logarithm on the right-hand side is always defined, we need to restrict the admissible values of $k$ to the interval $(-R_1, \infty)$. Inter-collision times are assumed to be independent, so for a sum $T$ of inter-collision times such as (1), we get the log of the Laplace transform of $T$ as
\begin{equation}
\lambda(k) = -\ln \langle \text{e}^{-kT} \rangle = -\ln \bigg \langle \prod_{n=1}^{\mathcal{N}} \text{e}^{-kt_n} \bigg \rangle = -\ln \prod_{n=1}^{\mathcal{N}}\langle \text{e}^{-kt_n} \rangle = \sum_{n=1}^{\mathcal{N}} \ln\bigg(1 + \frac{k}{R_n}\bigg)
\end{equation}
The expression on the right-hand side of (21) appears as equation (14) in MW2016. This is differentiable with respect to $k$ for all $k \in (-R_1, \infty)$, so by the G\"artner-Ellis Theorem of large-deviation analysis, discussed in \cite{Touchette}, the rate function in (16) is given by
\begin{equation}
J(\tau) = \sup_k \{\lambda(k) - k\tau\}
\end{equation}
The expression on the right-hand side of (22) is referred to as the \emph{Fenchel-Legendre transform} of $\lambda(k)$. 

Note that the expression in (22) is the negative of the expression usually given in large deviation theory. This is because the outer function in the formulation of the log of the Laplace transform in (20) above is the negative of the expression used, e.g., in \cite{Touchette}, and this formulation also has the parameter $k$ negated. The alternative formulation would be 
\begin{equation*}
\lambda_n(k) = \ln \langle \text{e}^{kt_n} \rangle = \ln\bigg(\int_0^{\infty} d t_n \text{e}^{kt_n} R_n \text{e}^{-R_n t_n}\bigg) = -\ln\bigg(1 - \frac{k}{R_n}\bigg)
\end{equation*}
and the associated entropy function for this would then be the more conventional
\begin{equation*}
J(\tau) = \sup_k \{k\tau - \lambda(k)\}
\end{equation*}
In this case, the admissible values of $k$ would be those in the interval $(-\infty, R_1)$. Essentially, the signs of $\lambda(k)$ and $k$ have been negated in the formulation in MW2016, as this is more convenient for the analysis.

Differentiating the bracketed expression in (22) with respect to $k$ and setting equal to zero we obtain
\begin{equation}
\lambda^{\prime}(k) - \tau = 0
\end{equation}
But
\begin{equation}
\lambda^{\prime}(k) = \sum_{n=1}^{\mathcal{N}} \frac{\big(\frac{1}{R_n}\big)}{1 + \frac{k}{R_n}} = \sum_{n=1}^{\mathcal{N}} \frac{1}{R_n + k}
\end{equation}
Putting this in (23) we get
\begin{equation}
\tau =  \sum_{n=1}^{\mathcal{N}} \frac{1}{R_n + k}
\end{equation}
This implicitly gives an optimal value of $k$ as a function of $\tau$, say $k^{*}(\tau)$. If this could be obtained explicitly, inserting it into the right-hand side of (22) would give the required rate function $J(\tau)$:
\begin{equation}
J(\tau) = \lambda(k^{*}(\tau)) -  k^{*}(\tau) \cdot \tau
\end{equation} 
However, in practice it is not possible to obtain an exact expression for $J(\tau)$ in this way. Instead, MW2016 obtains aymptotic expressions for $J(\tau)$ which are valid for small $\tau$. This is achieved by reparameterising $J(\tau)$ using a scaled variable $\kappa$, defined as
\begin{equation}
\kappa = \frac{k^{*}}{R_1}
\end{equation} 
From (7) and (8) we get, as $\mathcal{N} \rightarrow \infty$,
\begin{equation}
\tau = \frac{\bar{T}}{\langle T \rangle} = \frac{R_1 \bar{T}}{\zeta(\gamma)}
\end{equation}
Using (5), (27) and (28) in (25) we get 
\begin{equation}
\tau(\kappa) = \frac{R_1}{\zeta(\gamma)} \sum_{n=1}^{\infty} \frac{1}{R_1 n^{\gamma} + R_1 \kappa} = \frac{1}{\zeta(\gamma)} \sum_{n=1}^{\infty} \frac{1}{\kappa + n^{\gamma}}
\end{equation}
The expression on the right-hand side of (29) appears in equation (18) in MW2016. 

We now write (21) in reparameterised form as $\mathcal{N} \rightarrow \infty$ as
\begin{equation}
S(\kappa) = \sum_{n=1}^{\infty} \ln \bigg(1 + \frac{R_1 \kappa}{R_1 n^{\gamma}}\bigg) = \sum_{n=1}^{\infty} \ln(1 + \kappa n^{-\gamma})
\end{equation}
We also write 
\begin{equation}
k^{*}(\tau) \cdot \tau = \frac{R_1 \kappa \zeta(\gamma) \tau}{R_1} = \zeta(\gamma) \kappa \tau
\end{equation}
Then we re-express (26) in the form
\begin{equation}
J(\kappa)= S(\kappa) - \zeta(\gamma) \kappa \tau(\kappa)
\end{equation}
This appears as equation (19) in MW2016. 

Following the approach outlined in MW2016S, the next stage in the development is to obtain an asymptotic expression for the sum in (30) as $\kappa \rightarrow \infty$. (It is assumed from now on that $R_1 = 1$, to simplify the expressions). Upon differentiating this, we will then obtain a corresponding asymptotic expression for $\tau$ in accordance with (29), which can be inverted to obtain a small $\tau$ approximation for $\kappa$, and subsequently a small $\tau$ approximation for $J(k)$. Finally, this will be used in (19) to obtan an expression for $\tau^{*}$ in terms of $\ln\mathcal{N}^{*}$. 

We find that
\begin{equation}
S \sim S_0 - \frac{1}{2} \ln(\kappa) - \gamma C + O(\kappa^{-1})
\end{equation}
where 
\begin{equation*}
S_0 = \int_0^{\infty} dn \ln(1 + \kappa n^{-\gamma})
\end{equation*}
and where $C$ is a constant with numerical value
\begin{equation}
C \approx 0.91896611
\end{equation}
Equation (33) appears as equation (10) in MW2016S. 

From (29), we see that differentiating $S(\kappa)$ in (2.30) gives a time $T(\kappa) = \zeta(\gamma) \tau(\kappa)$. The corresponding asymptotic expression for $T(\kappa)$ is obtained by differentiating (33). We get
\begin{equation}
\frac{\partial S}{\partial \kappa} \sim \int_0^{\infty} dn \frac{1}{\kappa} \bigg(\frac{1}{1 + \frac{n^{\gamma}}{\kappa}}\bigg) - \frac{1}{2\kappa}
\end{equation}
In the integral in (35), make the change of variable $x = \frac{n^{\gamma}}{\kappa}$, so that $(\kappa x)^{1/\gamma} = n$. Then
\begin{equation*}
dn = \frac{1}{\gamma} (\kappa x)^{\frac{1}{\gamma} -1} \cdot \kappa dx = \frac{1}{\gamma} \kappa^{-\frac{\gamma -1}{\gamma}} \cdot x^{-\frac{\gamma -1}{\gamma}} \cdot \kappa dx
\end{equation*}
The integral in (35) then becomes 
\begin{equation*}
\kappa^{-\frac{\gamma -1}{\gamma}} \cdot \frac{1}{\gamma} \int_0^{\infty} dx \frac{x^{-\frac{\gamma -1}{\gamma}}}{1+x}
\end{equation*}
so we get 
\begin{equation}
\frac{\partial S}{\partial \kappa} = T(\kappa) \sim A(\gamma) \cdot \kappa^{-\frac{\gamma -1}{\gamma}} - \frac{1}{2\kappa}
\end{equation}
with 
\begin{equation}
A(\gamma) \equiv \frac{1}{\gamma} \int_0^{\infty} dx \frac{x^{-\frac{\gamma -1}{\gamma}}}{1+x}
\end{equation}
(MW2016 notes that the integral in (37) is the beta function $B\big(1-\frac{1}{\gamma}, \frac{1}{\gamma}\big)$). Integrating the first term on the right-hand side of (36), which is $\frac{\partial S_0}{\partial \kappa}$, gives
\begin{equation}
S_0 = \gamma A(\gamma) \kappa^{\frac{1}{\gamma}}
\end{equation}
Substituting for $S_0$ in (33) with the expression in (38) gives
\begin{equation}
S \sim \gamma A(\gamma) \kappa^{\frac{1}{\gamma}} - \frac{1}{2} \ln(\kappa) - \gamma C + O(\kappa^{-1})
\end{equation} 
This is equation (21) in MW2016. 

To obtain an explicit expression for the rate function $J(\tau)$, we must now invert (36) to express the parameter $\kappa$ in terms of the time $\bar{T}$, and hence in terms of the dimensionless time given by $\tau = \frac{\bar{T}}{\zeta(\gamma)}$. We find that
\begin{equation}
\kappa = b\tau^{-\frac{\gamma}{\gamma-1}}[1 - c \tau^{\frac{1}{\gamma-1}}]
\end{equation}
with 
\begin{equation}
b \equiv \bigg(\frac{A}{\zeta(\gamma)}\bigg)^{\frac{\gamma}{\gamma-1}}
\end{equation}
and
\begin{equation}
c \equiv \frac{\gamma}{2(\gamma-1)A(\gamma)}b^{-\frac{1}{\gamma}}
\end{equation}
Now, using (36) and (39), the rate function is given by 
\begin{equation*}
J(\tau) = S(\kappa^{*}) - T(\kappa^{*})\kappa^{*}
\end{equation*}
\begin{equation*}
= \gamma A(\gamma) \kappa^{\frac{1}{\gamma}} - \frac{1}{2} \ln(\kappa) - \gamma C -A \kappa^{\frac{1}{\gamma}} + \frac{1}{2}
\end{equation*}
\begin{equation}
= (\gamma - 1) A\kappa^{\frac{1}{\gamma}} - \frac{1}{2} \ln(\kappa) - \bigg(\gamma C - \frac{1}{2}\bigg)
\end{equation}
Using (40) in (43) we get
\begin{equation}
J(\tau) = (\gamma - 1) A b^{\frac{1}{\gamma}} \tau^{-\frac{1}{\gamma-1}} + \frac{1}{2} \frac{\gamma}{\gamma - 1} \ln(\tau) + D
\end{equation}
where $D$ is another constant. 

We are now finally in a position to obtain a key result given in equation (27) in MW2016. From (44) we see that to leading order we have
\begin{equation*}
J(\tau^{*}) \propto [\tau^{*}]^{-\frac{1}{\gamma-1}}
\end{equation*}
so from (19) we obtain
\begin{equation}
\tau^{*} \propto [\ln \mathcal{N}^{*}]^{-(\gamma - 1)}
\end{equation}
Thus, the model predicts that the time scale required for runaway raindrop growth decreases as the number of collisions required for the onset of runaway growth increases. (Note that there is actually an error in equation (27) in MW2016. The power is incorrectly given as $-\frac{\gamma-1}{\gamma}$ there, instead of the correct $-(\gamma-1)$).

MW2016 does not provide direct numerical accuracy checks of the above asymptotic formulae against their corresponding exact large-deviation theory functions. Instead, MW2016 reports accuracy checks for a log-density function $\ln[P(\tau)]$, where $P(\tau)$ is defined in equation (22) of MW2016S as
\begin{equation*}
P(\tau) = \frac{1}{\sqrt{2 \pi J^{\prime \prime}(\tau)}}\exp[-J(\tau)]
\end{equation*}  
with
\begin{equation}
J^{\prime \prime}(\tau) = \frac{\gamma-1}{\gamma}A b^{-\frac{2\gamma-1}{\gamma-1}}\tau^{\frac{2\gamma-1}{\gamma-1}}
\end{equation}
According to the numerical results reported in MW2016, this function of the above asymptotic formulae provides a reasonable approximation of the same function of the corresponding exact large-deviation theory functions. 

It is an easy matter, however, to also directly check the accuracy of, say, equation (40) above (which is equation (17) in MW2016S), equation (43) above (which is equation (19) in MW2016S) and equation (46) above (which is equation (21) in MW2016S) against their exact large-deviation theory counterparts. Such checks reveal that the asymptotic formulae (17) to (21) in MW2016S are actually rather inaccurate when they are individually compared with corresponding exact function values. For example, Figures 1 to 3 are superimposed plots of asymptotic formulae (17), (19) and (21) in MW2016S against exact function values with $N=10^4$ and $\gamma=2$, showing that the asymptotic formulae are inaccurate. The results for the case $\gamma=4/3$ were even worse (not shown here), and there was no improvement when $N$ was raised to $10^5$.

The log-density function which features in the numerical work in MW2016 is a combination of two of the inaccurate asymptotic formulae, namely (19) and (21) in MW2016S, and it is clear that combining these formulae in this particular way produced a masking of their individual inaccuracy in MW2016.  
\begin{figure}
\caption{Superimposed plots of $\kappa(\tau)$ for $N=10^4$, $\gamma=2$.}
\centering
\includegraphics[width=\textwidth]{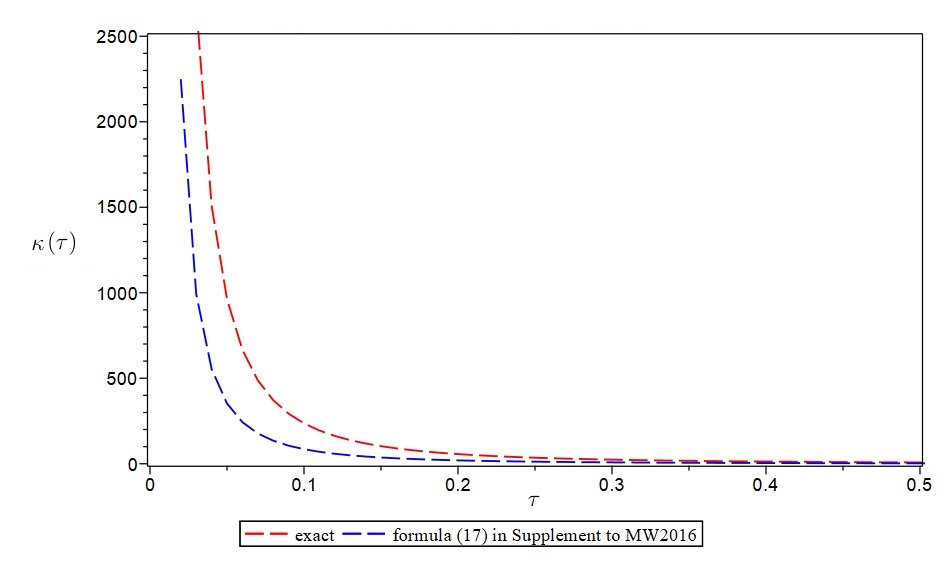}
\end{figure}
\begin{figure}
\caption{Superimposed plots of $J(\tau)$ for $N=10^4$, $\gamma=2$.}
\centering
\includegraphics[width=\textwidth]{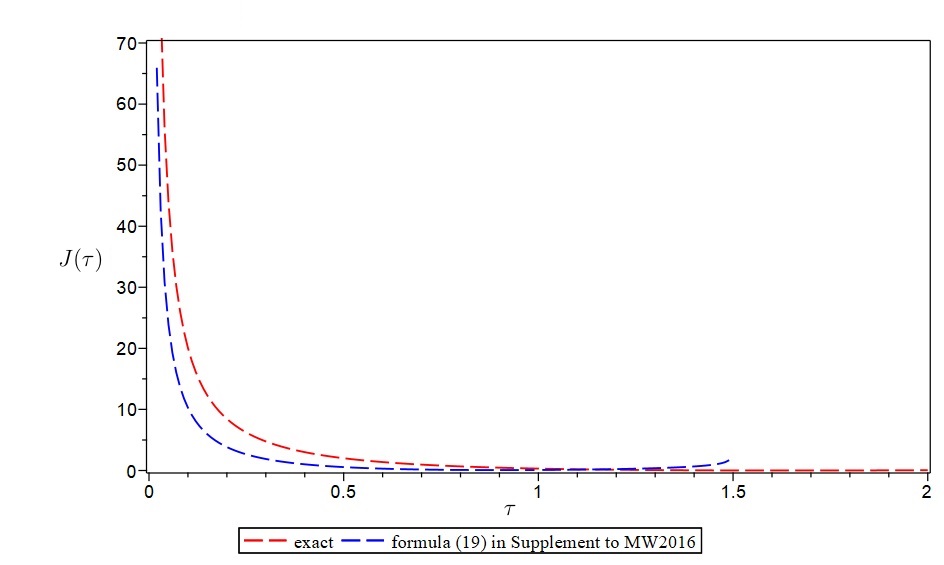}
\end{figure}
\begin{figure}
\caption{Superimposed plots of $J^{\prime \prime}(\tau)$ for $N=10^4$, $\gamma=2$.}
\centering
\includegraphics[width=\textwidth]{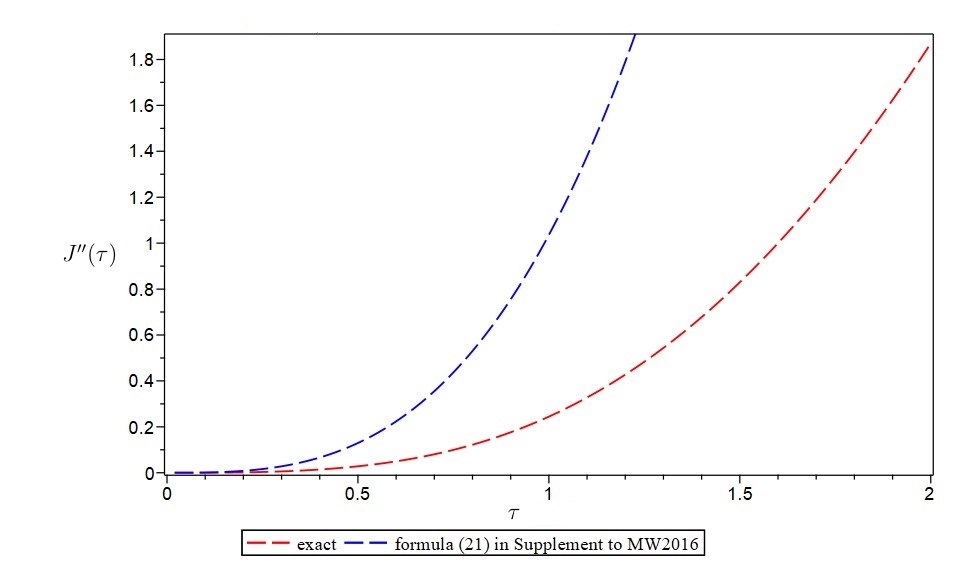}
\end{figure}
\newpage
\section{Accurate analytical formulae for the large-deviation functions}
We now employ a non-asymptotic approximation scheme which produces extremely accurate analytical expressions for the cumulant generating functions and entropy functions in MW2016. We will not need the transformations of $k$ into $\kappa$ and $\bar{T}$ into $\tau$ that were employed in MW2016, and since no asymptotic approximation methods are used, the analytical expressions set out in this section will be valid for any relevant values of $N$, $k$ and $\bar{T}$.
\subsection{The general forms of the analytical formulae}
Below we will focus on the cases of most interest to us, $\gamma=2$ and $\gamma=4/3$, but, for general $\gamma>1$, the non-asymptotic analytical expressions preserve the leading order power law form of the asymptotic formulas for $S(k^{\ast})$ in equation (21) in MW2016, and for $J(k^{\ast})$ in equation (19) of the Supplement to MW2016. The non-asymptotic analytical expressions developed here all have the form 
\begin{equation*}
\lambda(N, k, \gamma) = A(N, k, \gamma) \cdot k^{1/\gamma} + B(N, k, \gamma) 
\end{equation*}
\begin{equation}
+\frac{1}{2}\bigg(\ln\bigg(1+\frac{k}{N^{\gamma}}\bigg)+\ln(1+k)\bigg) -\bigg(\frac{1}{6}\bigg)\bigg(\frac{1}{2}\bigg)\bigg(\frac{\gamma k}{N(N^{\gamma}+k)}-\frac{\gamma k}{1+k}\bigg)
\end{equation}
for the cumulant generating function, and
\begin{equation*}
J(N, k, \gamma) = A(N, k, \gamma) \cdot k^{1/\gamma} + B(N, k, \gamma) + C(N, k, \gamma)
\end{equation*}
\begin{equation}
+\frac{1}{2}\bigg(\ln\bigg(1+\frac{k}{N^{\gamma}}\bigg)+\ln(1+k)\bigg) -\bigg(\frac{1}{6}\bigg)\bigg(\frac{1}{2}\bigg)\bigg(\frac{\gamma k}{N(N^{\gamma}+k)}-\frac{\gamma k}{1+k}\bigg)
\end{equation}
for the entropy function, where the functions $A$, $B$ and $C$ are to be obtained explicitly in exact form using the following formulae:
\begin{equation}
A(N, k, \gamma) = \int_{1/k}^{N^{\gamma}/k} dy \  \frac{y^{-\frac{\gamma-1}{\gamma}}}{1+y}
\end{equation}
\begin{equation}
B(N, k, \gamma) = N\cdot\ln\bigg(1+\frac{k}{N^{\gamma}}\bigg) - \ln(1+k)
\end{equation}
\begin{equation}
C(N, k, \gamma) = - k \cdot \frac{\partial \lambda(N, k, \gamma)}{\partial k}
\end{equation}
The integrand in the formula for $A(N, k, \gamma)$ in (49) has exactly the same form as the integrand in the formula for $A(\gamma)$ given in equation (22) in MW2016, and $A(N, k, \gamma)$ is therefore to be regarded as the non-asymptotic analogue of the latter. The last two terms in (47) and (48) come from the Euler-Maclaurin formula \cite{Apostol2} 
\begin{equation*}
\lambda(N, k; \gamma) = \int_1^N dn f(n) + \frac{1}{2} \big(f(1) + f(N)\big) 
\end{equation*}
\begin{equation}
- \sum_{m=1}^{\mathcal{M}} \frac{B_{2m}}{(2m)!} \big(f^{(2m-1)}(1) - f^{(2m-1)}(N)\big) + R_{\mathcal{M}+1}
\end{equation}
where $f(n) = \ln\big(1+\frac{k}{n^{\gamma}}\big)$, $B_{2m}$ are the even-indexed Bernoulli numbers, and $R_{\mathcal{M}+1}$ is a remainder term. The factor 1/6 that appears in the expressions is the second Bernoulli number, $B_2$, which is from the first term in the Bernoulli number series in the Euler-Maclaurin formula. 

In the case $\gamma=2$, the functions $A$, $B$ and $C$ are obtained explicitly as follows:
\newline
\begin{equation}
A(N, k, 2) = 2\cdot\bigg\{\arctan\bigg(\frac{N}{\sqrt{k}}\bigg)-\arctan\bigg(\frac{1}{\sqrt{k}}\bigg)\bigg\}
\end{equation}
\newline
\begin{equation}
B(N, k, 2) = N \cdot \ln\bigg(1 + \frac{k}{N^2}\bigg) - \ln(1+k)
\end{equation}
\newline
\begin{equation*}
C(N, k, 2) = -\bigg\{\arctan\bigg(\frac{N}{\sqrt{k}}\bigg)-\arctan\bigg(\frac{1}{\sqrt{k}}\bigg)\bigg\}\cdot k^{1/2} 
\end{equation*}
\begin{equation*}
+ \frac{k\cdot\bigg(\frac{1}{6}N - \frac{2}{3}N^4-\frac{1}{2}N^2\bigg)}{(N^2+k)^2\cdot(k+1)^2}+ \frac{k^2\cdot\bigg(\frac{1}{3}N - \frac{1}{2}N^4-\frac{7}{3}N^2-\frac{1}{2}\bigg)}{(N^2+k)^2\cdot(k+1)^2}
\end{equation*}
\begin{equation}
+ \frac{k^3\cdot\bigg(-\frac{5}{3}-\frac{3}{2}N^2 + \frac{1}{6}N\bigg)}{(N^2+k)^2\cdot(k+1)^2} - \frac{k^4}{(N^2+k)^2\cdot(k+1)^2}
\end{equation}

In the case $\gamma=4/3$, the exact functions $A$, $B$ and $C$ are as follows:
\newline
\begin{equation*}
A(N, k, 4/3) = \sqrt{2}\cdot\big\{\arctan(\sqrt{2}\cdot k^{1/4}-1)+\arctan(\sqrt{2}\cdot k^{1/4}+1)\big\}
\end{equation*}
\begin{equation*}
- \sqrt{2}\cdot\bigg\{\arctan\bigg(\sqrt{2}\cdot \frac{k^{1/4}}{N^{1/3}}-1\bigg)+\arctan\bigg(\sqrt{2}\cdot \frac{k^{1/4}}{N^{1/3}}+1\bigg)\bigg\}
\end{equation*}
\begin{equation*}
+\frac{1}{\sqrt{2}}\cdot\big\{\ln(\sqrt{2}\cdot k^{1/4}+1+\sqrt{k})-\ln(-\sqrt{2}\cdot k^{1/4}+1+\sqrt{k})\big\}
\end{equation*}
\begin{equation}
+ \frac{1}{\sqrt{2}}\cdot\bigg\{\ln\bigg(-\sqrt{2}\cdot \frac{k^{1/4}}{N^{1/3}}+1+\frac{\sqrt{k}}{N^{2/3}}\bigg)-\ln\bigg(\sqrt{2}\cdot \frac{k^{1/4}}{N^{1/3}}+1+\frac{\sqrt{k}}{N^{2/3}}\bigg)\bigg\}
\end{equation}
\newline
\begin{equation}
B(N, k, 4/3) = N \cdot \ln\bigg(1 + \frac{k}{N^{4/3}}\bigg) - \ln(1+k)
\end{equation}
\newline
\begin{equation*}
C(N, k, 4/3) = -\frac{3}{4} \cdot A(N, k, 4/3) \cdot k^{3/4}
\end{equation*}
\begin{equation*}
-\frac{((N^{4/3}-N)k-(N-1)k^2)}{{\scriptstyle (\sqrt{2}\cdot k^{1/4}-1-\sqrt{k})(\sqrt{2}\cdot k^{1/4}+1+\sqrt{k})(\sqrt{2}\cdot N^{1/3} k^{1/4}-N^{2/3}-\sqrt{k})(\sqrt{2}\cdot N^{1/3}k^{1/4}+N^{2/3}+\sqrt{k})}}
\end{equation*}
\begin{equation}
-\frac{(N+1/2)k}{N^{4/3}+k} + \frac{7k}{18(k+1)} + \frac{k^2}{9(k+1)^2}-\frac{k^2}{9N(N^{4/3}+k)^2}+\frac{k}{9N(N^{4/3}+k)}
\end{equation}
\newline
The derivation of the above results is described in detail in the next section. To implement the above formulae, we obtain values of $k^{\ast}(\bar{T})$ for any given values of $\bar{T}$ as the zeros of the following equation using the first derivative of the analytical expression for the cumulant generating function in (47) above:
\begin{equation}
\frac{\partial \lambda(N, k, \gamma)}{\partial k} - \bar{T} = 0
\end{equation}  
As the derivative on the left-hand side involves only simple arctan and log functions and powers of $N$ and $k$, this can be done extremely quickly for any desired number of values of $\bar{T}$ (e.g., using fsolve in MAPLE). The values of $k^{\ast}(\bar{T})$ are then substituted into the above analytical expressions, and also into the analytical expressions for the first and second derivatives of the cumulant generating function in (47) above. The saddlepoint density function $P(k^{\ast}(\bar{T}))$ is obtained as
\begin{equation}
P(k^{\ast}(\bar{T})) = \frac{\exp\big(-J(k^{\ast}(\bar{T}))\big)}{\sqrt{2\pi \big[-\frac{\partial^2 \lambda(N, k^{\ast}(\bar{T}), \gamma)}{\partial k^2}\big]}}
\end{equation} 
\subsection{Derivation using the Euler-Maclaurin formula}
The cumulant generating function in equation (14) in MW2016 is 
\begin{equation}
\lambda(N, k, \gamma) = \sum_{n=1}^N \ln\bigg(1 + \frac{k}{n^{\gamma}}\bigg)
\end{equation}
The first step in the derivation of the non-asymptotic formulae set out in the previous subsection is to approximate this using the Euler-Maclaurin formula given in equation (52) in the previous subsection. Using only the first term in the Bernoulli number series, the approximation is obtained as
\newpage
\begin{equation*}
\lambda(N, k; \gamma) = \int_1^N dn \ln\bigg(1 + \frac{k}{n^{\gamma}}\bigg) 
\end{equation*}
\begin{equation}
+ \frac{1}{2}\bigg(\ln\bigg(1 + \frac{k}{N^{\gamma}}\bigg)+\ln\big(1 + k\big)\bigg) - \bigg(\frac{1}{6}\bigg)\bigg(\frac{1}{2}\bigg)\bigg(\frac{\gamma k}{N(N^{\gamma}+k)}-\frac{\gamma k}{1+k}\bigg)
\end{equation} 
The next step is to manipulate the integral appearing in this approximation by making the change of variable $x=\frac{k}{n^{\gamma}}$. We then have
\begin{equation}
n = \bigg(\frac{k}{x}\bigg)^{1/\gamma} = k^{1/\gamma} x^{-1/\gamma}
\end{equation}
and
\begin{equation}
dn = -\frac{1}{\gamma} \  k^{1/\gamma} \ x^{-\frac{1+\gamma}{\gamma}} dx
\end{equation} 
When $n=1$, $x=k$, and when $n=N$, $x = \frac{k}{N^{\gamma}}$. The integral in (62) becomes
\begin{equation}
\frac{1}{\gamma} \  k^{1/\gamma} \int_{k/N^{\gamma}}^k dx \  x^{-\frac{1+\gamma}{\gamma}} \ln(1+x)
\end{equation}
Using integration by parts, we then obtain
\begin{equation*}
\int_1^N dn \ln\bigg(1 + \frac{k}{n^{\gamma}}\bigg) 
\end{equation*}
\begin{equation*}
= \frac{1}{\gamma} \  k^{1/\gamma} \int_{k/N^{\gamma}}^k dx \  x^{-\frac{1+\gamma}{\gamma}} \ln(1+x)
\end{equation*}
\begin{equation*}
= \frac{1}{\gamma} \  k^{1/\gamma} \bigg\{\bigg[-\gamma x^{-1/\gamma} \ln(1+x)\bigg]_{k/N^{\gamma}}^k + \int_{k/N^{\gamma}}^k dx \ \gamma x^{-1/\gamma} \bigg(\frac{1}{1+x}\bigg)\bigg\}
\end{equation*}
\begin{equation}
= k^{1/\gamma} \int_{k/N^{\gamma}}^k dx \ \frac{x^{-1/\gamma}}{1+x} + N\cdot\ln\bigg(1+\frac{k}{N^{\gamma}}\bigg) - \ln(1+k) 
\end{equation}
Finally, make the change of variable $x=\frac{1}{y}$ in the final integral in (66), to obtain
\begin{equation}
\int_1^N dn \ln\bigg(1 + \frac{k}{n^{\gamma}}\bigg) = k^{1/\gamma} \int_{1/k}^{N^{\gamma}/k} dy \  \frac{y^{-\frac{\gamma-1}{\gamma}}}{1+y}  + N\cdot\ln\bigg(1+\frac{k}{N^{\gamma}}\bigg) - \ln(1+k) 
\end{equation}
Substituting (67) into (62) then gives the non-asymptotic analytical expression for $\lambda(N, k, \gamma)$ given in equation (47) in the previous subsection, with $A(N, k, \gamma)$ and $B(N, k, \gamma)$ as defined in (50) and (51). 

To obtain the explicit expression for $A(N, k, 2)$ given in equation (53) in the previous subsection, consider the indefinite form of the final integral in (66), with $\gamma=2$. Making the change of variable $u = \sqrt{x}$ in this integral gives
\begin{equation}
\int dx \ \frac{x^{-1/2}}{1+x} = 2\int du \ \bigg(\frac{1}{1+u^2}\bigg) = 2 \arctan(u) 
\end{equation}
Substituting back $u = \sqrt{x}$ into this result and evaluating as a definite integral with limits $k/N^2$ and $k$ gives
\begin{equation*}
A(N, k, 2) = 2\cdot\bigg\{\arctan\big(\sqrt{k}\big)-\arctan\bigg(\frac{\sqrt{k}}{N}\bigg)\bigg\} 
\end{equation*}
\begin{equation}
= 2\cdot\bigg\{\arctan\bigg(\frac{N}{\sqrt{k}}\bigg)-\arctan\bigg(\frac{1}{\sqrt{k}}\bigg)\bigg\}
\end{equation} 
where the last equality follows from the identity $\arctan(1/x) = \pi/2 - \arctan(x)$ for $x > 0$. To obtain the explicit expression for $A(N, k, 4/3)$ given in equation (56) in the previous section, consider again the indefinite form of the final integral in (66), this time with $\gamma=4/3$:
\begin{equation}
\int dx \ \frac{x^{-3/4}}{1+x} = \int dx \ \frac{x^{1/4}}{x(1+x)} 
\end{equation}
Making the change of variable $u = x^{1/4}$ gives
\begin{equation}
\int dx \ \frac{x^{1/4}}{x(1+x)} = 4\int \frac{du}{1 + u^4}  
\end{equation}
We now expand the integrand on the right-hand side in partial fractions using
\begin{equation}
1 + u^4 = (u^2 + 1)^2 - 2 u^2 = \big((u^2 + 1) + \sqrt{2}u\big) \big((u^2 + 1) - \sqrt{2}u\big)
\end{equation}
Therefore we seek to expand in partial fractions as follows:
\begin{equation}
\frac{1}{1+u^4} = \frac{A}{(u^2 + 1) + \sqrt{2}u} + \frac{B}{(u^2 + 1) - \sqrt{2}u}
\end{equation}
Multiplying through by (72) gives
\begin{equation}
1 = A\big((u^2 + 1) - \sqrt{2}u\big) + B\big((u^2 + 1) + \sqrt{2}u\big)
\end{equation}
The only zero of $\big((u^2 + 1) + \sqrt{2}u\big)\big((u^2 + 1) - \sqrt{2}u\big)$ is given by $(u^2 + 1) = \sqrt{2}u$ which implies $1 = u(\sqrt{2}-u)$. Substituting into (74) we obtain
\begin{equation}
B = \frac{\sqrt{2}-u}{2\sqrt{2}}
\end{equation}
Using (74) again, we require $A$ such that 
\begin{equation}
1 = \big(A + B \big)(u^2 + 1) + \big(B - A \big) \sqrt{2}u
\end{equation}
Thus, we need $A$ such that $A + B = 1$ and $B - A = -\frac{u}{\sqrt{2}}$. Using (75), we deduce
\begin{equation}
A = \frac{\sqrt{2}+u}{2\sqrt{2}}
\end{equation}
The required partial fraction expansion is then (73), with $A$ and $B$ as given in (75) and (77). The integral in (71) then becomes
\begin{equation}
4\int \frac{du}{1 + u^4} = \sqrt{2} \int du \ \frac{u + \sqrt{2}}{(u^2+1)+\sqrt{2}u} + \sqrt{2} \int du \ \frac{\sqrt{2}-u}{(u^2+1)-\sqrt{2}u}
\end{equation}
The first integral on the right-hand side of (78) can be expanded as
\begin{equation}
\int du \ \frac{\sqrt{2}u}{(u^2+1)+\sqrt{2}u} + \int du \ \frac{2}{(u^2+1)+\sqrt{2}u}
\end{equation}
To deal with this, observe that
\begin{equation}
\frac{d\ln(u^2 + \sqrt{2}u + 1)}{du} = \frac{2u + \sqrt{2}}{u^2 + \sqrt{2} + 1} = \frac{\sqrt{2}(\sqrt{2}u+1)}{u^2 + \sqrt{2} + 1}
\end{equation}
Therefore
\begin{equation}
\frac{\sqrt{2}u}{(u^2+1)+\sqrt{2}u} = \frac{1}{\sqrt{2}} \frac{d\ln(u^2 + \sqrt{2}u + 1)}{du}  -  \frac{1}{u^2 + \sqrt{2} + 1}
\end{equation}
Integrating we get
\begin{equation}
\int du \ \frac{\sqrt{2}u}{(u^2+1)+\sqrt{2}u} = \frac{1}{\sqrt{2}}\ln(u^2 + \sqrt{2}u + 1) - \int du \ \frac{1}{(u^2+1)+\sqrt{2}u}
\end{equation}
Therefore (79) reduces to 
\begin{equation}
\frac{1}{\sqrt{2}}\ln(u^2 + \sqrt{2}u + 1) + \int du \ \frac{1}{(u^2+1)+\sqrt{2}u}
\end{equation}
But 
\begin{equation}
\int du \ \frac{1}{(u^2+1)+\sqrt{2}u} = 2\int du \ \frac{1}{1+(\sqrt{2}u+1)^2} = \sqrt{2} \arctan(\sqrt{2}u + 1)
\end{equation}
Therefore (79) is
\begin{equation*}
\frac{1}{\sqrt{2}}\ln(u^2 + \sqrt{2}u + 1) +  \sqrt{2} \arctan(\sqrt{2}u + 1)
\end{equation*}
so the first integral on the right-hand side of (78) is
\begin{equation}
\sqrt{2} \int du \ \frac{u + \sqrt{2}}{(u^2+1)+\sqrt{2}u} = \frac{1}{\sqrt{2}}\ln(u^2 + \sqrt{2}u + 1) +  \sqrt{2} \arctan(\sqrt{2}u + 1)
\end{equation}
Analogously, the second integral on the right-hand side of (78) is obtained as 
\begin{equation}
\sqrt{2} \int du \ \frac{\sqrt{2}-u}{(u^2+1)-\sqrt{2}u} = -\frac{1}{\sqrt{2}}\ln(u^2 - \sqrt{2}u + 1) +  \sqrt{2} \arctan(\sqrt{2}u - 1)
\end{equation}
We conclude that
\begin{equation*}
4\int \frac{du}{1 + u^4} = \frac{1}{\sqrt{2}}\ln(u^2 + \sqrt{2}u + 1)  - \frac{1}{\sqrt{2}}\ln(u^2 - \sqrt{2}u + 1)
\end{equation*}
\begin{equation}
\qquad \qquad \qquad + \sqrt{2} \arctan(\sqrt{2}u + 1) + \sqrt{2} \arctan(\sqrt{2}u - 1)
\end{equation}
Finally, substituting back $u = x^{1/4}$ into this result and evaluating as a definite integral with limits $k/N^{4/3}$ and $k$ gives $A(N, k, 4/3)$ as defined in equation (56) in the previous subsection. 
\subsection{Accuracy of the analytical formulae}
In a checking process analogous to the one for MW2016 described in Section 2, stringent tests were carried out of the non-asymptotic analytical expressions set out in Section 3.1 for the cases $\gamma=2$ and $\gamma=4/3$, with varying values for $N$ up to $N=10^4$, by comparing them with sets of fine-grained numerical data obtained from the corresponding exact functions. Superimposed plots of the exact data and the non-asymptotic formulae were produced to visually assess accuracy. All the superimposed plots were, without exception, virtually indistinguishable, as exemplified by Figures 4 to 7 below, showing that the non-asymptotic analytical expressions proposed in this note are indeed extremely accurate. 

\begin{figure}
\caption{Superimposed plots of $J(\bar{T})$ for $N=10^4$, $\gamma=2$.}
\centering
\includegraphics[width=\textwidth]{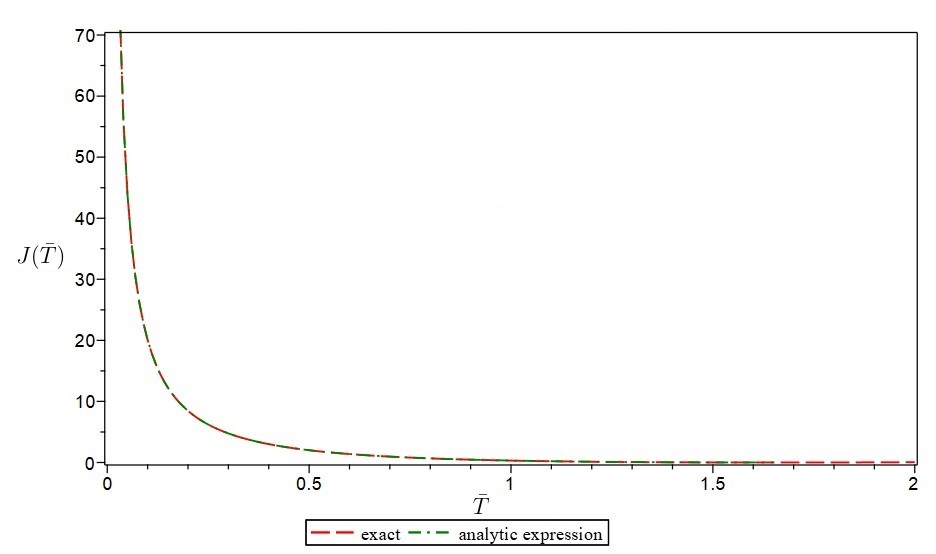}
\end{figure}
\begin{figure}
\caption{Superimposed plots of $\ln P[\bar{T}]$ for $N=10^4$, $\gamma=2$.}
\centering
\includegraphics[width=\textwidth]{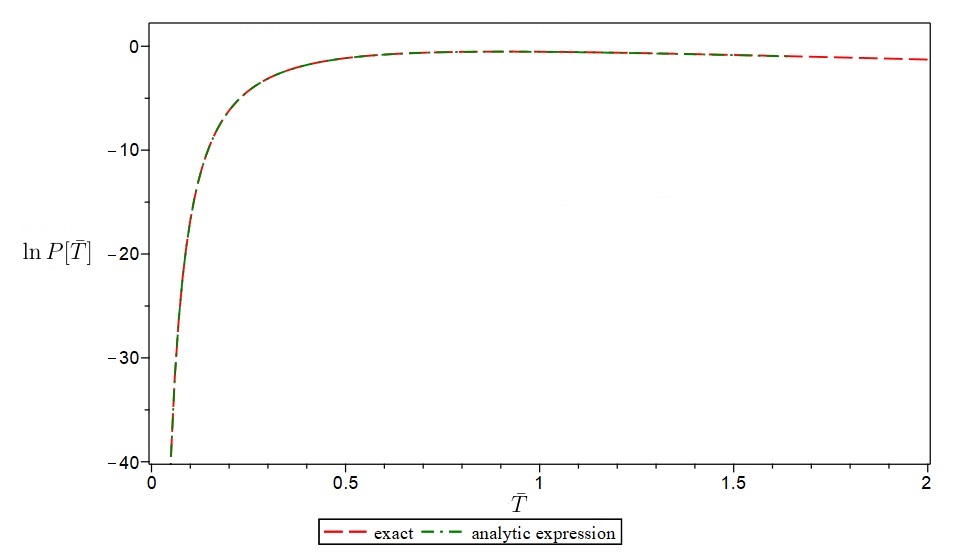}
\end{figure}
\begin{figure}
\caption{Superimposed plots of $J(\bar{T})$ for $N=10^4$, $\gamma=4/3$.}
\centering
\includegraphics[width=\textwidth]{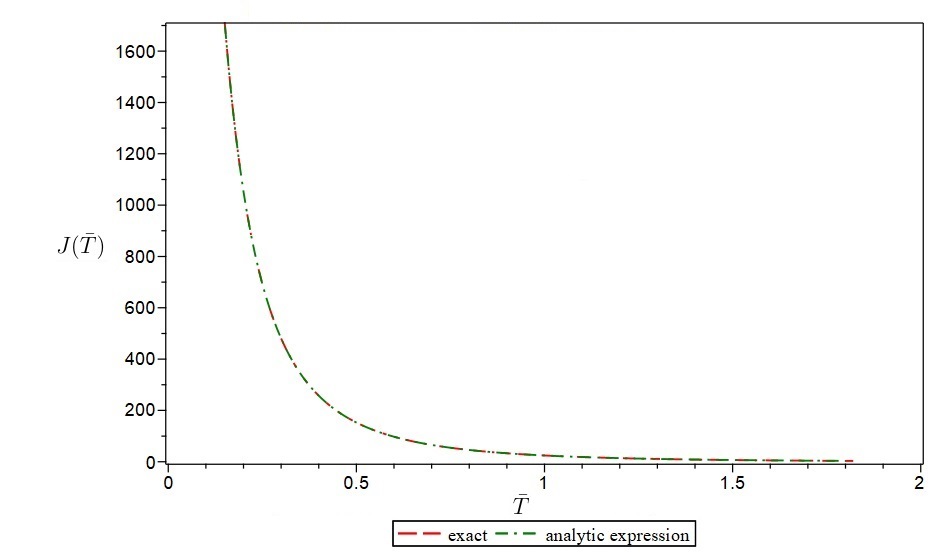}
\end{figure}
\begin{figure}
\caption{Superimposed plots of $\ln P[\bar{T}]$ for $N=10^4$, $\gamma=4/3$.}
\centering
\includegraphics[width=\textwidth]{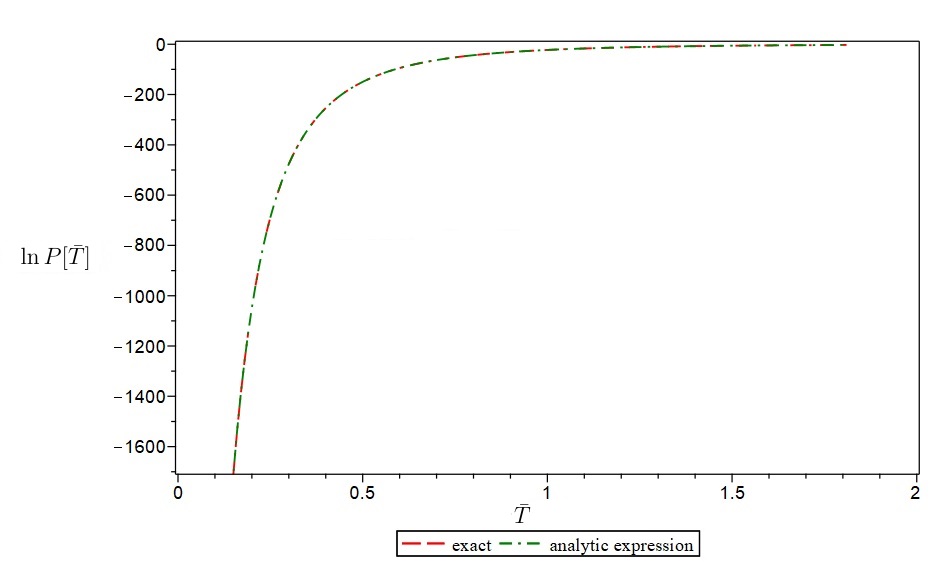}
\end{figure}

\newpage
\section{Discussion}
This note has developed highly accurate \emph{non-asymptotic} analogues of the asymptotic large-deviation approximation formulae in MW2016 which preserve some key features of the latter, particularly the \emph{power law} forms of the cumulant generating function and entropy function in MW2016. The aim was to avoid making any simplifying asymptotic assumptions involving $N$, $k$ or $\bar{T}$, as we may need to allow these key variables to range over intermediate values when applying the theory to new problems. Asymptotic formulae derived by assuming very large values of $N$ and/or $k$ (the latter corresponds to assuming small values of $\bar{T}$) are unlikely to be directly applicable in these wider contexts. 

The accuracy checks reported in Section 2 led to additional numerical investigations which indicated that there may be two principal sources of asymptotic approximation errors in MW2016. One of these is the sequence of approximations leading to the asymptotic formula $A(\gamma)$ in equation (15) in MW2016S, presented as equation (22) in MW2016. This asymptotic formula is independent of $N$ and $k$, and is therefore treated as a constant in the cumulant generating function in equation (21) in MW2016. In deriving the non-asymptotic analytical expressions in the present paper, the asymptotic assumptions leading to $A(\gamma)$ were avoided, resulting in a non-asymptotic analogue of this formula which is allowed to vary with $N$ and $k$. Since this version is allowed to vary with $k$ in particular, its derivatives will contribute to the first and second derivatives of the cumulant generating function with respect to $k$. These derivatives seem to play significant roles in accurately approximating the entropy function and the saddlepoint probability density function. Since $A(\gamma)$ appears as a constant in the cumulant generating function in MW2016, the derivatives that are contributing to the greater accuracy in the non-asymptotic formulation in the present note are simply missing in the asymptotic formulae for the entropy and saddlepoint density in MW2016. Correcting this problem immediately produced improvements in approximation accuracy in the non-asymptotic approach in the present note.

The other main source of approximation error in MW2016 is likely to be that $A(\gamma)$ is further used to express $k$ as an explicit function of $\bar{T}$ in equations (17) and (18) in MW2016S. The constancy of $A(\gamma)$ with respect to $k$ is actually a necessary simplification for this, but it leads to even more approximation error being introduced into the expressions involving the entropy function (because $k(\bar{T})$ is crucial in defining the entropy). This issue is simply sidestepped in the present note by obtaining values of $k(\bar{T})$ from the implicit function defined by the first derivative of the (non-asymptotic) analytical approximation of the cumulant generating function. Numerically, this is just as easy as using an explicit function for $k$ and, again, this simple step dramatically improved the approximation accuracy in the non-asymptotic approach herein. 

The accurate non-asymptotic analytical expressions in this note were obtained largely by avoiding the above two problems, and also by using the first term from the Bernoulli number series in the Euler-Maclaurin formula introduced in Section 3. Since the (exact) cumulant generating function in equation (14) in MW2016 is so well-behaved, it was only necessary to include this first term in the series to achieve almost perfect approximation accuracy. 

\section*{Acknowledgement}
I have benefitted from numerous discussions with Michael Wilkinson about a range of issues arising from his work on the large-deviation analysis of rapid-onset rainfall.  

\centerline{}

\end{document}